\documentclass[manuscript]{aastex61}
\usepackage[T1]{fontenc}

\usepackage{bm}        
\usepackage{amssymb, amsmath}   
\usepackage{cancel}
\usepackage{color}

\newcommand\aastex{AAS\TeX}

\DeclareMathAlphabet{\mathsfit}{\encodingdefault}{\sfdefault}{m}{sl}
\SetMathAlphabet{\mathsfit}{bold}{\encodingdefault}{\sfdefault}{bx}{sl}
\newcommand{\vect}[1]{\bm{#1}}

\newcommand{\red}{\textcolor{red}}
\DeclareMathOperator{\sech}{sech}
\newcommand{\tensorsym}[1]{\bm{\mathsfit{#1}}}

\received{\today}
\revised{***}
\accepted{***}
\submitjournal{ApJ}

\shorttitle{\aastex\ Compression Acceleration in Magnetic Reconnection}
\shortauthors{Li et al.}

\begin{document}

\title{Large-scale Compression Acceleration during Magnetic Reconnection in a Low-$\beta$ Plasma}

\correspondingauthor{Xiaocan Li}
\email{xiaocanli@lanl.gov}

\author[0000-0001-5278-8029]{Xiaocan Li}
\affil{Los Alamos National Laboratory, Los Alamos, NM 87545, USA}

\author{Fan Guo}
\affiliation{Los Alamos National Laboratory, Los Alamos, NM 87545, USA}
\affiliation{New Mexico Consortium, Los Alamos, NM 87544, USA}


\author{Hui Li}
\affiliation{Los Alamos National Laboratory, Los Alamos, NM 87545, USA}

\author{Shengtai Li}
\affiliation{Los Alamos National Laboratory, Los Alamos, NM 87545, USA}

\begin{abstract}
In solar flares and other astrophysical systems, a major challenge for solving
particle acceleration problem associated with magnetic reconnection is the
enormous scale separation between kinetic scales and observed reconnection
scale. Because of this, it has been difficult to draw any definite conclusions by just
using kinetic simulations. Particle acceleration model that solves energetic
particle transport equation can capture the main acceleration physics found in
kinetic simulations, and thus provide a practical way to make observable
predictions and directly compare model results with observations. Here we study
compression particle acceleration in magnetic reconnection by
solving Parker (diffusion-advection) transport equation using velocity and magnetic fields from
two-dimensional high-Lundquist-number magnetohydrodynamics (MHD) simulations of
a low-$\beta$ reconnection layer. We show that the compressible reconnection layer can give
significant particle acceleration, leading to the formation of power-law particle energy
distributions. We analyze the acceleration rate and find that the acceleration in
the reconnection layer is a mixture of first-order and second-order Fermi processes.
When including a guide field, we find the spectrum becomes steeper
and both the power-law cutoff energy and maximum particle energy decrease as plasma
becomes less compressible. This model produces a 2D particle
distribution that one can use to generate radiation map and directly compare
with solar flare observations. This provides a framework to explain particle
acceleration at large-scale astrophysical reconnection sites, such as solar
flares.
\end{abstract}


\section{Introduction}
\label{sec:intro}
Energy conversion and particle acceleration in strongly magnetized plasmas are
important processes that hold the key for understanding many explosive solar
and astrophysical high-energy phenomena~\citep{Zweibel2009Magnetic, Lin2011Energy}.
Magnetic reconnection is a major mechanism that drives the release of magnetic energy and
nonthermal particle acceleration by reorganizing the topology and connectivity of
magnetic field lines~\citep{Fu2013Energetic, Fu2017Intermittent}.
One of the best examples for magnetic reconnection and the associated particle
acceleration is solar flares. Observations have suggested that magnetic
reconnection converts 10\% to 50\% of the magnetic energy (up to $\sim 10^{33}$ ergs)
into plasma kinetic energy within $1-10$ minutes. During the process, a large amount
of electrons in the flare region ($>10^{36}$ electrons) are accelerated into a power-law energy
spectrum $f(\varepsilon) \propto \varepsilon^{-s}$ with spectral index from
$s \sim 3$ to more than $s=9$
with a medium about $5$~\citep{Lin1976Nonthermal, Krucker2010Measure, Oka2013Kappa,
Oka2015Electron, Effenberger2017Hard}. The acceleration of ions in a flare region
can be as efficient as that of electrons. This is suggested by \textit{RHESSI}'s
observation on the correlation between electron-generated hard $X$-ray flux and
ion-generated $\gamma$-ray flux~\citep{Shih2009RHESSI}. In-situ solar energetic
particle (SEP) observation has also shown that the electron and ion spectra often
resemble power-law distributions~\citep{Mason2012Inter}. How such efficient
particle acceleration occurs over a large-scale reconnection region remains an
important unsolved problem in reconnection study. 

During solar flares, large-scale magnetic reconnection is in the weakly
collisional (high Lundquist number) regime and is likely to have magnetic
structures with a range of spatial scales. One attractive scenario emerged in
the past decade is the plasmoid-dominated reconnection, where a hierarchy of
plasmoids develop in a macroscopic reconnection layer
\citep{Shibata2001Plasmoid,Loureiro2007Instability, Bhattacharjee2009Fast,
Comisso2016General} and
naturally bring the current sheet from the macroscopic scale to kinetic scale
\citep{Daughton2009Transition,Ji2011Phase}. It is therefore important to study
particle acceleration in magnetic reconnection using a multi-scale approach.
For magnetic reconnection at kinetic scales, kinetic simulations provide first-principle
description for particle acceleration, but the domain size is
limited due to the demanding computational expense. The standard approach to
study particle acceleration on large scales is to solve energetic particle
transport equation~\citep[e.g.][]{Parker1965Passage, Zank2014Particle}, but this has not
been applied in reconnection study until recently (see below for a more detailed
discussion). Instead, test-particle simulations have been widely used to study
particle acceleration during reconnection on large scales. Below
we review the previous theories and numerical simulations on particle
acceleration in magnetic reconnection. 

Particle-in-cell (PIC) kinetic simulation has been popular in modeling
particle acceleration during magnetic reconnection, as it includes the full range
of plasma physics. Previous kinetic simulations have extensively studied several
acceleration mechanisms, such as direct acceleration close to the reconnection
$X$-point~\citep{Hoshino2001Suprathermal, Drake2005Production, Fu2006Process,
Oka2010Electron, Egedal2012Large, Egedal2015Double,Wang2016Mechanisms},
Fermi-type acceleration in contracting magnetic
islands~\citep{Drake2006Electron, Oka2010Electron}, acceleration in
island-merging regions~\citep{Oka2010Electron, Liu2011Particle, Drake2013Power,
Nalewajko2015Distribution}, and acceleration at reconnection
front~\citep{Fu2011Fermi, Fu2012Pitch, Liu2017Rapid, Liu2017Explaining, Xu2018Electron}.
By summing over particle guiding-center motions,
several recent studies have identified curvature drift along the motional
electric field as the major particle acceleration
mechanism~\citep{Dahlin2014Mechanisms, Guo2014Formation, Guo2015Particle,
Li2015Nonthermally, Li2017Particle} in the weak guide-field case. However,
because of the enormous scale separation between kinetic scales (ion skin depth
$\sim 10-100$ m) and scale of the observed reconnection region ($\sim 10^7$ m),
it has been difficult to draw any definite conclusion and compare solar flare
observations with the modeling results. To overcome this major difficulty and
solve particle acceleration problem in solar flare reconnection, one has to come
up with a description for the acceleration of particles in macroscopic fluid scale.

Test-particle simulations are widely used in studying particle acceleration
during solar flares. Both full particle orbits and the particle guiding-center
motions have been calculated in background electric and magnetic fields provided
by MHD simulations. Under the guiding-center approximation, one can solve
particle motions in realistic scales by removing the high-frequency
gyromotions. Test-particle method usually generates hard power-law energy
spectra~\citep{Onofri2006Stochastic, Gordovskyy2010ParticleB,
Gordovskyy2010ParticleA, Zhou2015Electron, Zhou2016Electron} that can extend to
tens of keV for electrons and tens of MeV for protons but may be too hard to
explain the observations (power-law index for electrons $1<s<2$). Acceleration
due to parallel electric field is usually the dominant particle acceleration
mechanism found in the these simulations. This is likely due to the large
anomalous resistivity and coarse grids used in these simulations, resulting in
much broader current layers and much larger resistive electric field than that
in real systems. Furthermore, the large anomalous resistivity is not supported by
current 3D PIC simulations of reconnection layers~\citep{Roytershteyn2012Influence,
Liu2013Bifurcated, Le2018Drift}. One can avoid this problem by ignoring the
parallel electric field completely~\citep{Zhou2015Electron, Birn2017Can}, leading
to particle energy spectra that are close to solar flare observations. But this
method still does not take into account the effect of wave-particle interaction that
scatters particles and changes the acceleration processes.

The standard approach to solve large-scale particle acceleration and transport
problem is to use the energetic particle transport theory, which has been widely
used in studying shock acceleration and cosmic ray transport. The primary
acceleration mechanism is due to adiabatic compression and is included in the
Parker transport equation~\citep{Parker1965Passage,Blandford1987Particle}.
Various other acceleration mechanisms (e.g. fluid shear and fluid acceleration)
could also be included in the transport theory~\citep{Earl1988Cosmic,
Zank2014Transport}. It is worthwhile noting that the acceleration due to
curvature drift and gradient drift found to be important in earlier kinetic simulations
has also been included in the transport theory~\citep{Jones1990Generalized,LeRoux2009Time}.
Several studies have attempted to develop similar transport theories (or reduced
kinetic equations) for studying particle acceleration during
reconnection~\citep{Drake2006Electron, Drake2013Power, Egedal2013Review,
Zank2014Transport, LeRoux2015Kinetic, Montag2017Impact}. These studies include
previously studied particle acceleration mechanisms, such as parallel reconnection
electric field and contracting and merging magnetic islands. While some of the
studies assume that the reconnection layer is incompressible and only consider
incompressible effects~\citep[e.g.][]{Drake2006Electron, Drake2013Power}, other
recent studies emphasized both compressible and incompressible
effects~\citep{Zank2014Particle, LeRoux2015Kinetic, Montag2017Impact}. Recently, for the
first time,~\citet{Li2018Roles} used fully kinetic simulations to show that
compression energization dominates the acceleration of high-energy particles in
reconnection with a weak guide field ($<20\%$ of the reconnecting component),
and the compression and shear effects are comparable when the guide field is
moderate ($\sim 0.5$ times of the reconnecting magnetic field component).
Meanwhile, some recent MHD simulations also suggest that the reconnection layer
is compressible especially when the plasma $\beta$ is low and the guide field is
weak~\citep{Birn2012Role, Provornikova2016Plasma}. These simulation results
suggest that one may study particle acceleration in a large-scale solar flare
reconnection site using the transport theory.

~\citet{Drury2012First} considered reconnection acceleration by assuming
the reconnection region as a black box with a certain compression ratio $r$
between the upstream and downstream regions. He found that compression acceleration
leads to a power-law spectrum $f(p) \propto p^{-\chi}$  and the spectral index
depends on the compression ratio in a similar way as in diffusive shock acceleration
$\chi = -3r/(r-1)$. For nonrelativistic particles, the spectral index $s$ for energy
spectrum $f(\varepsilon)$ is related to $\chi$ by $s = (\chi -1)/2$. As discussed above,
reconnection layer in the weakly collisional
regime may have magnetic structures in various scales. It is worthwhile studying
whether power-law energy spectrum can still develop and how the spectral features
depend on key plasma parameters of the reconnection layer. The goal of this paper
is to study large-scale compression acceleration during magnetic reconnection in
the plasmoid-dominated regime.

In this paper, we solve Parker (diffusion-advection) transport equation using the 
background velocity
and magnetic fields from high-Lundquist-number MHD simulations of a low-$\beta$
reconnection layer. We assume that electrons and protons are already energetic
and can interact with the background magnetic fluctuation existed in the
reconnection region. In Section~\ref{sec:methods}, we describe the MHD simulations
and stochastic integration method for solving the Parker transport equation.
In Section~\ref{sec:res}, we present our simulation results. We show that particles
are significantly accelerated by the compression reconnection layer in the
plasmoid-dominated regime. The acceleration leads to formation of power-law
energy distribution for both electrons and protons. The power-law index, cutoff energy
and the maximum energy depend on the guide-field strength and the diffusion
model. This model also produces 2D particle distribution that one can use to
generate radiation map and directly compare with observations. This provides a
framework to explain particle acceleration at large-scale reconnection sites,
such as solar flares. In
Section~\ref{sec:con}, we discuss the conclusions and the implications based on
our simulation results.

\section{Numerical Methods}
\label{sec:methods}
\subsection{MHD Simulations}
\label{subsec:mhd}
We carry out simulations of magnetic reconnection using the Athena MHD code~\citep{Stone2008Athena}.
We use a third-order piecewise parabolic reconstruction, the
Harten-Lax-van Leer Discontinuities (HLLD) Riemann solver, the MUSCL-Hancock (VL)
integrator, and the constrained transport (CT) algorithm to ensure the
divergence-free state of the magnetic field. The code solves the resistive MHD equations
\begin{gather*}
  \frac{\partial\rho}{\partial t} + \nabla\cdot(\rho\vect{v}) = 0, \\
  \frac{\partial(\rho\vect{v})}{\partial t} + \nabla\cdot\left[\rho\vect{v}\vect{v} +
  \left(p+\frac{\vect{B}\cdot\vect{B}}{2}\right)\tensorsym{I}-\vect{B}\vect{B}\right] = 0, \\
  \frac{\partial e}{\partial t} + \nabla\cdot\left[\left(e + p +
  \frac{\vect{B}\cdot\vect{B}}{2}\right)\vect{v} - \vect{B}(\vect{B}\cdot\vect{v})\right]
  = \nabla\cdot(\vect{B}\times\eta\vect{j}), \\
  \frac{\partial\vect{B}}{\partial t} - \nabla\times(\vect{v}\times\vect{B}) =
  \eta\nabla^2\vect{B},
\end{gather*}
where
\begin{equation*}
  e = \frac{p}{\gamma-1} + \frac{\rho\vect{v}\cdot\vect{v}}{2} +
  \frac{\vect{B}\cdot\vect{B}}{2},\quad \vect{j}=\nabla\times\vect{B},
\end{equation*}
where $\rho$ is the mass density, $\vect{v}$ is the velocity, $e$ is the total energy
density, $\vect{B}$ is the magnetic field, $p$ is the gas pressure, $\gamma$ (=5/3)
is the adiabatic index, $\vect{j}$ is the current density, and $\eta$ is the resistivity.
Unless specified otherwise, we normalize the simulations by $L_0=5000$ km (the simulation
box size is $10^4$ km) and $v_A=1000$ km s$^{-1}$, which are the typical parameters of
the reconnection site of a solar flare. We assume the normalized magnetic field
$B_0=50$ G and particle density is $1.2\times 10^{10} \text{cm}^{-3}$.
We choose $\eta=10^{-5}$ and the same box sizes $L_x=L_y=2$ in all simulations,
resulting a Lundquist number $S=L_yv_A/(2\eta)=10^5$. The simulation box is
$x\in[0, 2]$ and $y\in[0, 2]$. The simulations start from two current sheets with
\begin{align}
  \vect{B} = & B_0\left[\tanh\left(\frac{x-x_1}{\lambda}\right)-
  \tanh\left(\frac{x-x_2}{\lambda}\right)\right]\hat{y} - B_0\hat{y}+ \nonumber \\
  & B_0\left[\sqrt{\sech^2\left(\frac{x-x_1}{\lambda}\right) +
    \frac{B_{g}^2}{B_0^2}} + 
  \sqrt{\sech^2\left(\frac{x-x_2}{\lambda}\right) + 
\frac{B_{g}^2}{B_0^2}} - \frac{B_{g}}{B_0}\right]\hat{z},
\end{align}
where $B_0=1.0$ is the strength of the reconnecting magnetic field, $B_g$ is the
strength of the guide field, $x_1=0.5$ and $x_2=1.5$ are the $x$-positions of the
current sheets, and $\lambda=0.005$ is the half-thickness of the current sheet.
The grid sizes are $n_x\times n_y=8192\times4096$, so we can resolve the initial
current sheet by at least 10 cells.
We employ an initial magnetic flux perturbation to speed up the reconnection onset.
\begin{align}
  \psi_z(x,y) = \psi_0 B_0
  \left[\cos\left(\frac{2\pi (x-x_1)}{L_x}\right) -
  \cos\left(\frac{2\pi (x-x_2)}{L_x}\right)\right]
  \cos\left(\frac{2\pi y}{L_y}\right),
\end{align}
where $\psi_0$ is the amplitude of the perturbation. Initially the total pressure
(gas pressure + magnetic pressure) is uniform in the simulation box. We choose
$\psi_0=10^{-4}$ so that the initial density variation is under 2.6\%.  The
initial plasma $\beta=2p/B^2\approx 0.1$. We choose periodic boundary
conditions along both $x$ and $y$ directions. We perform 4 simulations with
$B_g=0$, 0.2, 0.5 and 1.0. The initial plasma density $\rho_0\approx 1.0$,
$\sqrt{1.04}$, $\sqrt{1.25}$, $\sqrt{2}$, so the resulting Alfv\'en speed
$v_A=\sqrt{(B_0^2+B_g^2)/\rho_0}$ in the reconnection inflow region $\approx1.0$
for all four runs. Note that the
Lundquist number in the simulations is much smaller than the realistic number
calculated from Coulomb collision. Previous numerical simulations have shown that
the reconnection rate becomes a few percent of the Alfv\'en speed and independent
of the Lundquist number when $S \gtrsim 10^4$~\citep[e.g.][]{Bhattacharjee2009Fast,
Huang2010Scaling}.

\subsection{Solving Parker Transport Equation}
\label{subsec:parker}
We then solve Parker's transport equation
\begin{equation}
  \frac{\partial f}{\partial t} + (\vect{v}+\vect{v}_d)\cdot\nabla f
  - \frac{1}{3}\nabla\cdot\vect{v}\frac{\partial f}{\partial\ln p}
  = \nabla\cdot(\vect{\kappa}\nabla f) + Q,
\end{equation}
where $f(x_i, p, t)$ is the particle distribution function as a function of the
particle position $x_i$,  momentum $p$ (isotropic momentum assumed),
and time $t$; $\vect{\kappa}$ is the spatial
diffusion coefficient tensor, $\vect{v}$ is the bulk plasma velocity, and $Q$ is
the source. Note that the particle drift $\vect{v}_d$ is out of the simulation
plane and is not considered here.
\red{
  We assume an isotropic particle distribution here based on earlier results of
  kinetic simulations~\citep{Li2018Roles}, where we showed that the anisotropy
  is weak for high-energy electrons and becomes
  even weaker as the simulation evolves. This indicates that plasma waves (or
  turbulence in 3D) can scatter particles to isotropize the particle distribution.
}
The diffusion coefficient tensor is given by
\begin{equation}
  \kappa_{ij} = \kappa_\perp\delta_{ij} -
  \frac{(\kappa_\perp-\kappa_\parallel)B_iB_j}{B^2},
\end{equation}
where $\kappa_\parallel$ and $\kappa_\perp$ are the parallel and perpendicular
diffusion coefficients. $\kappa_\parallel$ can be calculated from the quasilinear
theory~\citep{Jokipii1971Propagation}. Assuming that magnetic turbulence is
well-developed and has an isotropic power spectrum $P\sim k^{-5/3}$, the resulting
$\kappa_\parallel\sim p^{4/3}$ when the particle gyroradius is much smaller than
the correlation length of turbulence. In particular, we use the
following expression for $\kappa_\parallel$~\citep{Giacalone1999Transport},
\begin{equation}
  \kappa_\parallel(v) = \frac{3v^3}{20L_c\Omega_0^2\sigma^2}
  \csc\left(\frac{3\pi}{5}\right)\left[1+\frac{72}{7}
  \left(\frac{\Omega_0L_c}{v}\right)^{5/3}\right],
  \label{equ:kpara_qlt}
\end{equation}
where $v$ is the particle speed, $L_c$ is the turbulence correlation length,
$\Omega_0$ is the particle gyrofrequency, $\sigma^2=\left<\delta B^2\right>/B_0^2$
is the normalized wave variance of turbulence. The normalization of the diffusion
coefficient is then
$\kappa_0=L_0v_A=5\times10^{16}\text{ cm}^2\text{ s}^{-1}$, and the normalization
of time is $t_0=L_0/v_A = 5$ s. We assume that the correlation length
$L_c$ is equal to $\text{simulation box size} / 30 \approx 333\text{ km}$, which is the
largest eddy size in a reconnection-driven turbulence as shown by 3D MHD
simulations of magnetic reconnection~\citep{Huang2016Turbulent}. We assume the
average magnetic field $B_0=50$ G and $\sigma^2=1$. Then,
$\kappa_\parallel = 1.5\times 10^{14}\text{ cm}^2\text{ s}^{-1}$ for 10 keV
protons and $4.0\times 10^{14}\text{ cm}^2\text{ s}^{-1}$ for 1 keV electrons,
corresponding to $0.003\kappa_0$ and $0.008\kappa_0$ using simulation units.
Test-particle simulations have suggested that $\kappa_\perp/\kappa_\parallel$
is about 0.02-0.04 and is nearly independent of particle
energy~\citep{Giacalone1999Transport}. There are also observational evidence
suggesting $\kappa_\perp/\kappa_\parallel$ can be quite
large~\citep[e.g.,][]{Zhang2003Perp, Dwyer1997Perp}. Here we examine the effect of
$\kappa_\perp/\kappa_\parallel$ by adopting three different perpendicular diffusion 
$\kappa_\perp/\kappa_\parallel=0.01$, 0.05 and 1.0.

The Parker transport equation can be solved by integrating the stochastic differential
equation corresponding to the Fokker-Planck form of the transport
equation~\citep{Zhang1999Markov, Florinski2009Four, Pei2010General,Kong2017Acceleration}.
Neglecting the source term $Q$ in Equation (3) and assuming $F=fp^2$,
\begin{align}
  \frac{\partial F}{\partial t}
  & = -\nabla\cdot\left[(\nabla\cdot\vect{\kappa}+\vect{v})F\right] +
  \frac{\partial}{\partial p} \left[\frac{p}{3}\nabla\cdot\vect{v} F\right] +
  \nabla\cdot(\nabla\cdot(\vect{\kappa}F)),
\end{align}
which is equivalent to a system of stochastic differential equations (SDEs) of Ito type
\begin{equation}
  dX = (\nabla\cdot\vect{\kappa} + \vect{v})ds +
  \sum_\sigma\vect{\alpha}_\sigma dW_\sigma(s),\quad
  dp=-\frac{p}{3}(\nabla\cdot\vect{v})ds,
\end{equation}
where $\sum_\sigma\alpha_\sigma^\mu\alpha_\sigma^\nu = 2\kappa^{\mu\nu}$,
$dW$ is the normalized distributed random number with mean 0 and variance
$\sqrt{\Delta t}$,
and $\Delta t$ is the time step for stochastic integration. This corresponds to a Wiener
process. Numerical approximation is often-used for the Wiener process to replace
the normal distribution. We use a uniform distribution in $[-\sqrt{3}, \sqrt{3}]$
in the code. For a two-dimensional problem,
\begin{equation}
  \vect{\alpha}_1 =
  \begin{pmatrix}
    \sqrt{2\kappa_\perp} \\
    0
  \end{pmatrix}, \quad
  \vect{\alpha}_2 =
  \begin{pmatrix}
    0 \\
    \sqrt{2\kappa_\perp}
  \end{pmatrix}, \quad
  \vect{\alpha}_3 =
  \sqrt{2(\kappa_\parallel - \kappa_\perp)}
  \begin{pmatrix}
    B_x/B \\
    B_y/B
  \end{pmatrix}, \quad
\end{equation}

The parameters used at particle locations are
calculated from $v_x$, $v_y$, $B_x$, $B_y$, $\nabla\cdot\vect{v}$,
$\partial B_x/\partial x$, $\partial B_x/\partial y$,
$\partial B_y/\partial x$, $\partial B_y/\partial y$, which are all obtained from
the MHD simulations. We interpolate these parameters to the particle positions and
then calculate other required parameters:
\begin{align*}
  \frac{\partial\kappa_{xx}}{\partial x} & = \frac{\partial\kappa_\perp}{\partial x} - 
  \frac{\partial(\kappa_\perp-\kappa_\parallel)}{\partial x}\frac{B_x^2}{B^2} -
  2(\kappa_\perp-\kappa_\parallel)\frac{\frac{\partial B_x}{\partial x}B_xB-
  \frac{\partial B}{\partial x}B_x^2}{B^3}, \\
  \frac{\partial\kappa_{yy}}{\partial y} & = \frac{\partial\kappa_\perp}{\partial y} - 
  \frac{\partial(\kappa_\perp-\kappa_\parallel)}{\partial y}\frac{B_y^2}{B^2} -
  2(\kappa_\perp-\kappa_\parallel)\frac{\frac{\partial B_y}{\partial y}B_yB-
  \frac{\partial B}{\partial y}B_y^2}{B^3}, \\
  \frac{\partial\kappa_{xy}}{\partial x} & =
  -\frac{\partial(\kappa_\perp-\kappa_\parallel)}{\partial x}
  \frac{B_xB_y}{B^2} - (\kappa_\perp-\kappa_\parallel)
  \frac{\left(\frac{\partial B_x}{\partial x}B_y+
  B_x\frac{\partial B_y}{\partial x}\right)B -
  2B_xB_y\frac{\partial B}{\partial x}}{B^3}, \\
  \frac{\partial\kappa_{xy}}{\partial y} & =
  -\frac{\partial(\kappa_\perp-\kappa_\parallel)}{\partial y}
  \frac{B_xB_y}{B^2} - (\kappa_\perp-\kappa_\parallel)
  \frac{\left(\frac{\partial B_x}{\partial y}B_y+
  B_x\frac{\partial B_y}{\partial y}\right)B -
  2B_xB_y\frac{\partial B}{\partial y}}{B^3}, \\
  \frac{\partial B}{\partial x} & = \frac{1}{B}\left(B_x
  \frac{\partial B_x}{\partial x} + B_y\frac{\partial B_y}{\partial x}\right), \\
  \frac{\partial B}{\partial y} & =
  \frac{1}{B}\left(B_x\frac{\partial B_x}{\partial y} +
  B_y\frac{\partial B_y}{\partial y}\right).
\end{align*}
$\kappa_\parallel$ and $\kappa_\perp$ can be functions of $B_x$, $B_y$ and $B$,
so $\partial \kappa_\parallel/\partial x$, $\partial \kappa_\parallel/\partial y$,
$\partial \kappa_\perp/\partial x$, and $\partial \kappa_\perp/\partial y$ still
depend on the derivatives $\partial B_x/\partial x$, $\partial B_x/\partial y$,
$\partial B_y/\partial x$, $\partial B_y/\partial y$. The detailed expressions
depend on the diffusion model to choose.

In this work, we use a derivative-free Milstein method~\citep{Burrage2004Numerical}
to solve the stochastic differential equation. It is different from the usual method
due to one more term, which makes it become a higher-order method.
\begin{align}
  \frac{dX_t}{dt} & = f(X_t,t)dt + g(X_t,t)dW_t, \\
  X_{n+1} & = X_n + f_n h + g_n\Delta W_n +
  \frac{1}{2\sqrt{h}}[g(\bar{X}_n)-g_n][(\Delta W_n)^2-h], \\
  \bar{X}_n & = X_n + f_n h + g_n\sqrt{h}, \\
  \Delta W_n & = [W_{t+h}-W_t] \sim \sqrt{h}N(0,1),
\end{align}
where $X$ corresponds to spatial positions $x$, $y$ and particle momentum $p$
in our simulation. $f(X_t,t)$ is the deterministic term; $g(X_t,t)$ is the
probabilistic term; $h$ is the time step; $N(0,1)$ indicates a normal distribution,
which substituted with uniform distribution $[-\sqrt{3}, \sqrt{3}]$ in our
simulations to speed up the computation. For a 1D problem, the particle moves
a distance satisfying $l_x^2=\text{max}\left(\left<\Delta x\right>^2,
\left<\Delta x^2\right>\right)$~\citep{Strauss2017Hitch}, where
\begin{align}
  \left<\Delta x\right> = \left(v_x + \frac{d\kappa(x)}{dx}\right)\Delta t,
  \quad \left<\Delta x^2\right> = 2\kappa(x)\Delta t,
\end{align}
and $l_x$ should be much smaller than the spatial variation scale of the fields.
In this work, we assume $\left<\Delta x\right>^2 < \left<\Delta x^2\right>$ and
choose $\Delta t$ so that $l_x\ll\delta_x$, where $\delta_x$ is the grid size.
For our 2D problems, we choose the following criteria to determine the time step.
\begin{align}
  \Delta t_x & = \text{min}\left[\frac{\delta x}{80|v_x + \partial_x\kappa_{xx} +
  \partial_y\kappa_{xy}|},
  \frac{\left(\sqrt{2\kappa_\perp} + \sqrt{2(\kappa_\parallel - \kappa_\perp)}|B_x/B|\right)^2}
  {|v_x + \partial_x\kappa_{xx} + \partial_y\kappa_{xy}|^2}\right], \\
  \Delta t_y & = \text{min}\left[\frac{\delta y}{80|v_y + \partial_y\kappa_{yy} +
  \partial_x\kappa_{xy}|},
  \frac{\left(\sqrt{2\kappa_\perp} + \sqrt{2(\kappa_\parallel - \kappa_\perp)}|B_y/B|\right)^2}
  {|v_y + \partial_y\kappa_{yy} + \partial_x\kappa_{xy}|^2}\right],\\
  \Delta t & = \text{min}(\Delta t_x, \Delta t_y).
\end{align}

\section{Results}
\label{sec:res}
\subsection{Compression in a Reconnection Layer}
\begin{figure}[htbp]
  \centering
  \includegraphics[width=0.6\textwidth]{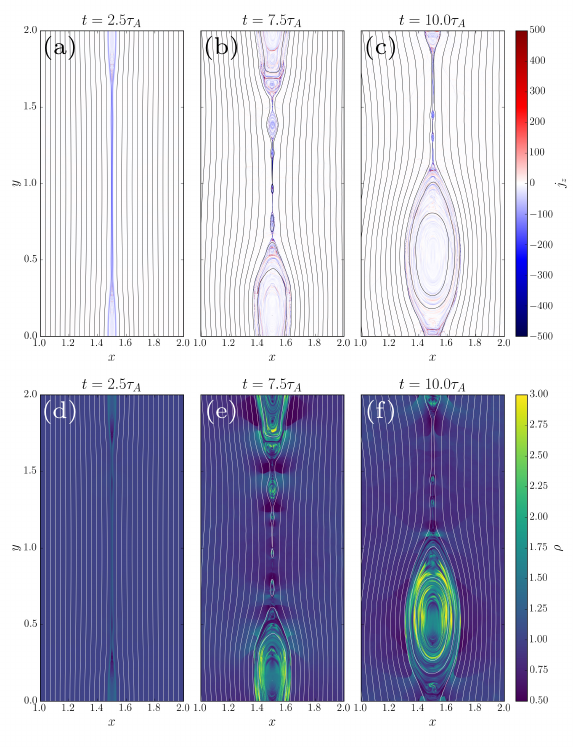}
  \caption{
    \label{fig:mhd_fields}
    The out-of-plane current density $j_z$ and plasma density $\rho$ at
    $t=2.5\tau_A$, $7.5\tau_A$, and $10\tau_A$ for half of the simulation box
    ($x=1.0-2.0$), where $\tau_A$ is the Alfv\'en
    crossing time $L_y/v_A$. The initial plasma density $\approx 1.0$.
  }
\end{figure}

As reconnection evolves, the current sheet becomes thinner and eventually unstable
to the plasmoid instability~\citep{Loureiro2007Instability, Bhattacharjee2009Fast,
Comisso2016General}. Figure~\ref{fig:mhd_fields} shows the
time evolution of the out-of-plane current density $j_z$ and plasma density $\rho$. 
At $t=2.5\tau_A$, where $\tau_A$ is the Alfv\'en crossing time $L_y/v_A$,
the current sheet just starts to break into magnetic islands
(Figure~\ref{fig:mhd_fields} (a) and (d)). These magnetic islands tend to contract
due to magnetic tension force and merge with each other to form larger islands
($t=7.5\tau_A$ and $10\tau_A$). Figure~\ref{fig:mhd_fields} (b) and (c) show that
new islands are continuously generated in the unstable current sheet. During these
processes, the maximum plasma density increases from 1.0 to 3.0 or higher
(Figure~\ref{fig:mhd_fields} (e) and (f)). The regions with enhanced density are
concentrated in magnetic islands, reconnection exhausts, and inflow regions around
the top and bottom sides of the magnetic islands. Due to the mass conservation
in the simulation domain, density
decreases in the inflow regions close to the reconnection layer and some regions
in the islands. Particles can be accelerated or decelerated when crossing these
regions. We expect that the net effect will be acceleration because on average,
the density increases as particles move from the inflow to the outflow regions.

\begin{figure}[htbp]
  \centering
  \includegraphics[width=\textwidth]{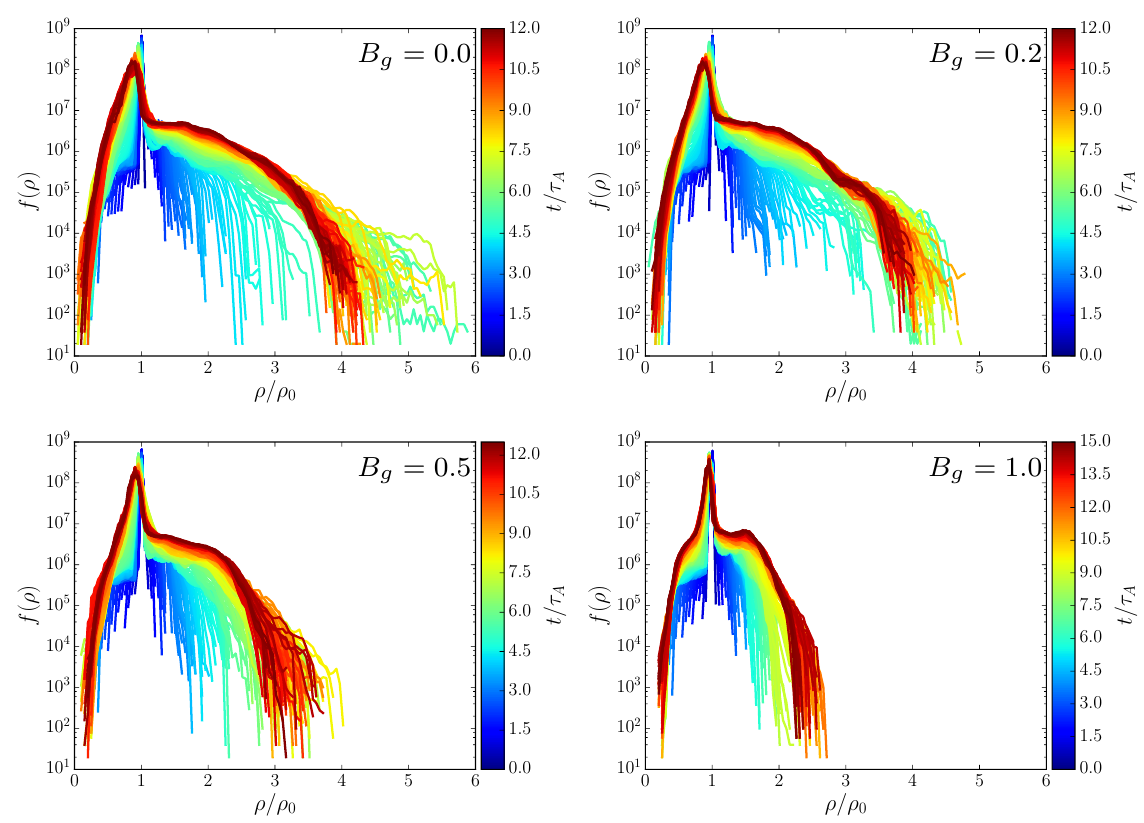}
  \caption{
    \label{fig:rho_dist}
    Time evolution of the density distributions for runs with different guide field.
    The plasma density is normalized by their initial values in each simulation.
    Time $t$ is normalized by the Alfv\'en crossing time $\tau_A=L_y/v_A$.
  }
\end{figure}

The enhanced plasma density suggests that the plasma in reconnection layer is
compressed. To further examine the plasma compressibility, Figure~\ref{fig:rho_dist}
shows the time evolution of the density distributions $f(\rho)$ for different runs.
Plasma density evolves to have a broad distribution from a nearly uniform value $\rho_0$
initially. The distributions constantly change as the simulation evolves, suggesting that
the reconnection layer is very dynamic. Take the run with $B_g=0$ for example,
$\rho/\rho_0$ reaches about 6 and then decreases to about 4, suggesting that the
compressed plasma in the reconnection layer can expand at late stage. Due to the mass
conservation, Figure~\ref{fig:rho_dist} shows significant distribution with $\rho/\rho_0<1$.
The guide field plays an important role in
controlling the plasma compressibility. As $B_g$ increases, the maximum density
decreases from about 6 when $B_g=0.0$ to 2.7 when $B_g=1.0$. This result is consistent
with previous MHD simulations~\citep{Birn2012Role, Provornikova2016Plasma}. Note that
$f(\rho)$ for $B_g=0.2$ is close to the case with $B_g=0$, indicating that a weak
guide field is not dynamically important here. This is because the magnetic
pressure from the guide field component 
is only 0.04 times that of the reconnecting component. The broad $f(\rho)$ and
the nonuniform spatial distribution of $\rho$ indicate that not all particles can
``see'' the entire density transition and that particle energy spectrum might not be a
simple function of the compression ratio as that predicted by diffusion-advection
analysis in a planar current sheet \citep{Drury2012First}.

\subsection{Particle Acceleration due to Compression: Constant Diffusion Coefficients}
\begin{figure}[htbp]
  \centering
  \includegraphics[width=0.5\textwidth]{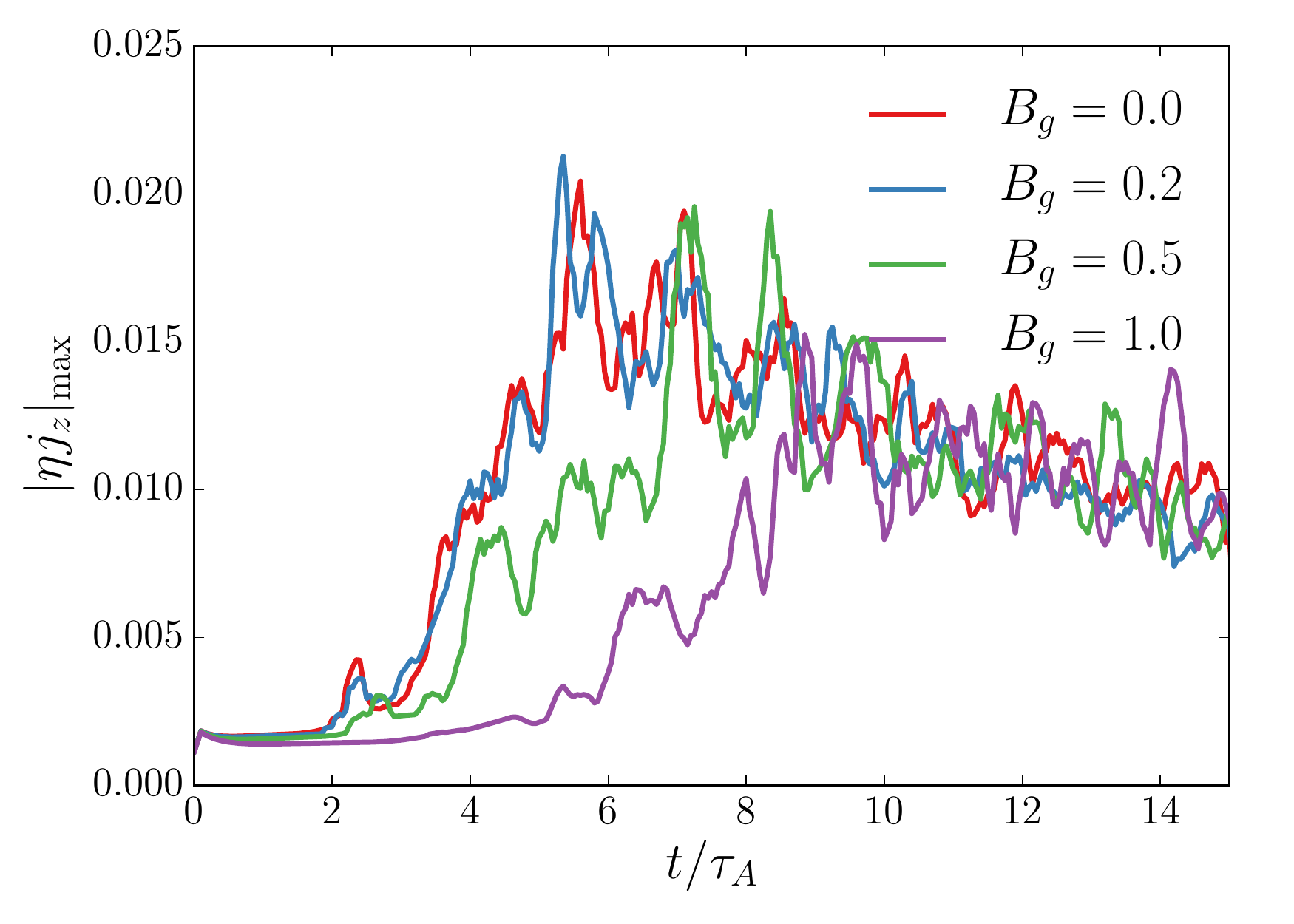}
  \caption{
    \label{fig:eta_jz}
    Time evolution of the maximum of $|\eta j_z|$ for different runs,
    where $\eta$ is the resistivity and $j_z$ is the out-of-plane current density.
  }
\end{figure}

The onset time for fast reconnection varies with the guide field.
Since we are mostly interested in the phase when the plasmoid instability is developed,
we start solving the acceleration of energetic particles by injecting pseudo
particles in the simulation when strong reconnection electric field emerges.
Figure~\ref{fig:eta_jz} shows the time evolution of the maximum value of
reconnection electric field $|\eta j_z|$, where $\eta$ is the resistivity.
$|\eta j_z|_\text{max}$ starts growing at different time as the guide field varies.
For runs with $B_g=0$ and 0.2, the rise time is almost the same. For runs with higher
$B_g$, it takes longer for $|\eta j_z|_\text{max}$ to grow. Based on this result, we
inject particles at $2\tau_A$ when $B_g=0$ or 0.2, at $2.5\tau_A$ when $B_g=0.5$,
and at $5\tau_A$ when $B_g=1.0$. For all the simulation cases, we continue to run
the simulation for $10\tau_A$ and solve the transport equation.

\begin{figure}[htbp]
  \centering
  \includegraphics[width=\textwidth]{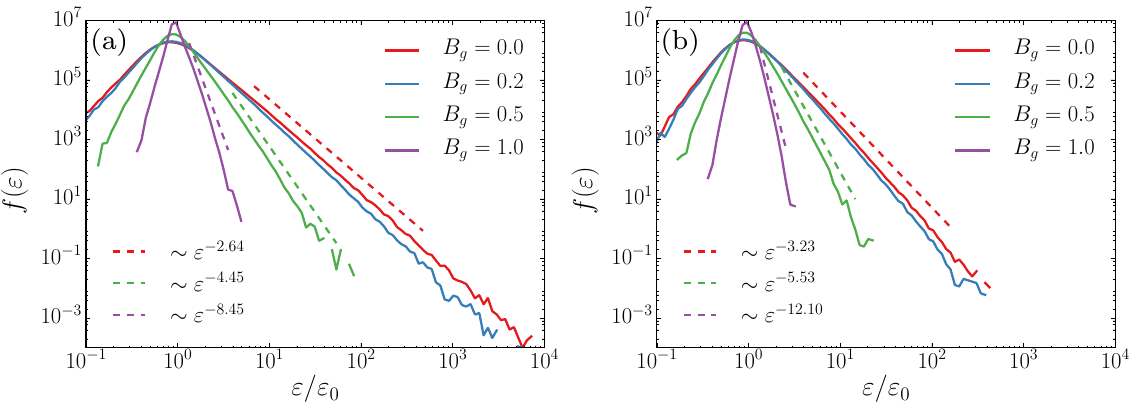}
  \caption{
    \label{fig:spect_k00038}
    Particle energy distributions for cases with constant diffusion coefficients.
    $p$ is particle momentum. $\varepsilon$ indicates particle energy and is
    normalized by the initial particle energy $\varepsilon_0$.
    The dashed lines indicate power-law fittings.
    For (a), we assume that the particles are protons with an initial energy
    10 keV and $\kappa_\parallel = \kappa_\perp = 0.003\kappa_0$.
    For (b), we assume that the particles are electrons with an initial energy
    1 keV and $\kappa_\parallel = \kappa_\perp = 0.008\kappa_0$.
  }
\end{figure}

We now discuss the results of energetic particle acceleration.
Figure~\ref{fig:spect_k00038} shows the final particle energy spectra when using
constant diffusion coefficients. We show two sets of simulations, one for protons
with an initial energy 10 keV and
$\kappa_\parallel = \kappa_\perp = 0.003\kappa_0$ (Figure~\ref{fig:spect_k00038} (a))
and the other for electrons with an initial energy 1 keV and
$\kappa_\parallel = \kappa_\perp = 0.008\kappa_0$ (Figure~\ref{fig:spect_k00038} (b)).
The eventual particle energy spectra resemble power-law distributions. When the guide field is weak, the
power-law distributions extend several orders of magnitude in energy. As the guide
field gets stronger, the power-law spectra become steeper and shorter, indicating
particle acceleration is more efficient in the reconnection with a
weaker guide field. The spectra are close to each other for cases with
$B_g=0$ and $B_g=0.2$. This is because the compressibility of the two cases
are close to each other (Figure~\ref{fig:rho_dist}). When $B_g$ increases to
1.0, particle spectrum becomes very steep with $f(\varepsilon)\sim\varepsilon^{-8.45}$
for protons and $f(\varepsilon)\sim\varepsilon^{-12.1}$ for electrons,
and the maximum energy is less than 10 times of the initial particle energy.
These results show that the guide-field strength is critical for particle
acceleration during magnetic reconnection. When the guide field is weak, the
plasma is strongly compressed in the reconnection layer, leading to an energy
spectrum harder than that of the strong guide field case. This trend for the
relation between the spectral index and the compressed plasma density is in agreement
with \citet{Drury2012First}, except that the spectral index also has a weak dependence
on the diffusion coefficient. 

\begin{figure}[htbp]
  \centering
  \includegraphics[width=\textwidth]{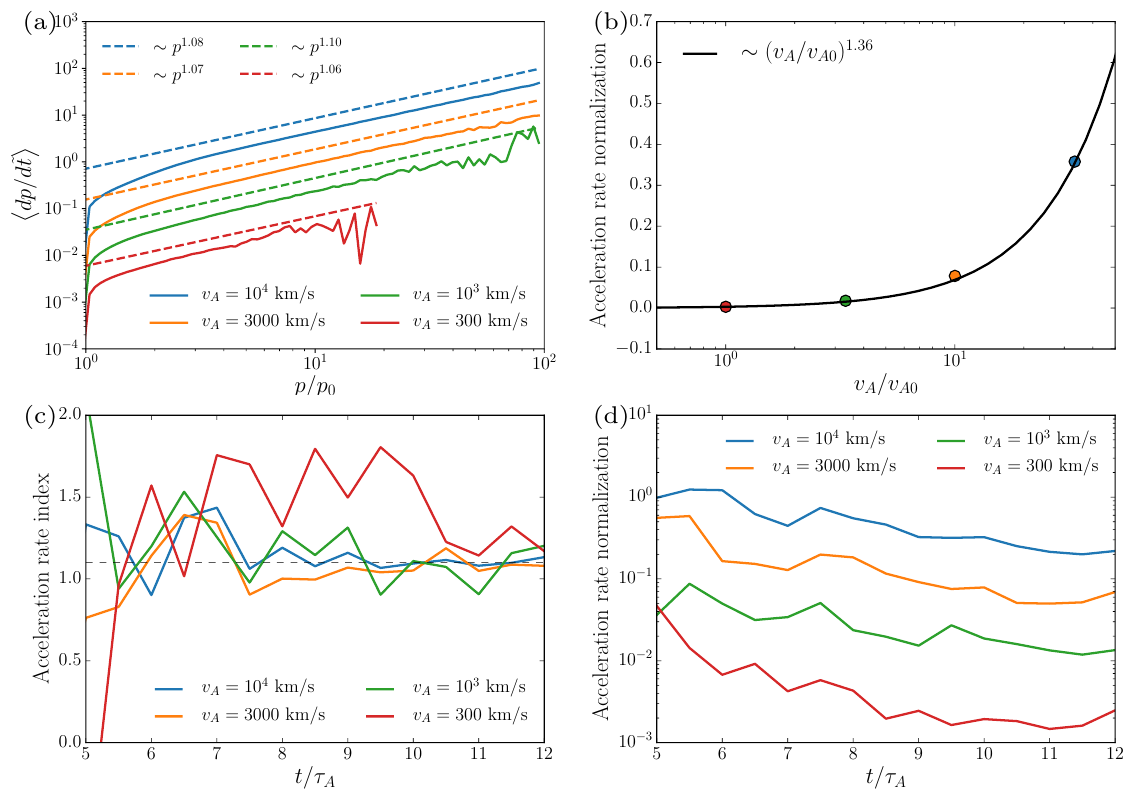}
  \caption{
    \label{fig:ene_scaling}
    Diagnostics on particle acceleration rate for simulations with constant
    $\kappa$. We wary the value of $\kappa$ by changing the Alfv\'en speed $v_A$
    and keeping
    the length scale $L_0$ constant. Here, we use the MHD run with $B_g=0$.
    (a) $\left<dp/d\tilde{t}\right>$ as a function of particle momentum.
    Note that we have normalized the simulation time $t$ in all runs with $L_0/v_{A0}$,
    which is $300$ km/s in our normalization, so $\tilde{t}=tv_{A0}/L_0$.
    We accumulate $dp/dt$ and particle number $n_p$ in each momentum bin
    every 0.0005$\tau_A$ from $2\tau_A$ to $12\tau_A$ and calculate
    $\left<dp/d\tilde{t}\right>=(\sum dp/dt)v_A/(n_pv_{A0})$.
    The solid lines are simulation data and the dashed lines are the power-law
    fittings $Cp^\alpha$, where $C$ is the acceleration rate normalization and
    $\alpha$ is the acceleration rate index. Note that the power-law fitting is
    shifted for better visualization. (b) The scaling of $C$ and hence
    $\left<dp/d\tilde{t}\right>$ with respect to the $v_A$.
    The four dots correspond to the four runs in (a). The black solid line is the
    power-law fitting. We normalize $v_A$ by $v_{A0}$.
    (c) Time evolution of the accelerate rate
    index starting at $t=5\tau_A$, when particles can be accelerated to fairly
    high energies. The black dashed line indicates an acceleration rate index 1.1.
    (d) Time evolution of the acceleration rate normalization $C$.}
\end{figure}

To examine the nature of particle acceleration in a reconnection layer,
we then study how the particle acceleration rate depends on the flow speed,
which is about the Alfv\'en speed $v_A$
in a reconnection layer. We add another three simulations with fixed
$\kappa_\parallel=\kappa_\perp= 1.5\times 10^{14} \text{cm}^2/\text{s}$
and $L_0 = 5000$ km but different $v_A$ from
$300$ km/s to $10^4$ km/s for the MHD run with $B_g=0$. The normalized $\kappa_\parallel$
and $\kappa_\perp$ then change from $0.01\kappa_0$ to $3\times10^{-4}\kappa_0$.
For each pseudo-particle, we calculate the acceleration rate $dp/dt = \Delta p / \Delta t$
for each short time interval $\Delta t =$ 0.0005$\tau_A$. Then, we statistically
calculate the acceleration rate for all particles in the system.
Figure~\ref{fig:ene_scaling} (a) shows the distributions of $\left<dp/d\tilde{t}\right>$,
averaged from $t=2\tau_A$ to $t=12\tau_A$, where we have normalized the simulation time $t$
with $L_0/v_{A0}$, and $v_{A0}$ is $300$ km/s in our normalization.
The measured acceleration rate is close to zero near the injected momentum
since most of injected pseudo-particles are outside of the reconnection layer in
the beginning. At higher energies, the acceleration rate becomes a power-law like
distribution as a function of momentum $\left<dp/d\tilde{t}\right> = C(p/p_0)^\alpha$. The
acceleration rate index $\alpha$ is 1.06 -- 1.10, which is expected as particles
gain energy through the compression term $-p\nabla\cdot\vect{v}/3$ in Parker
transport equation. Figure~\ref{fig:ene_scaling} (a) also shows that the acceleration
rate increases when the Alfv\'en speed gets larger. 
To further study the scaling of the acceleration
rate with respect to $v_A$, we fit $C$ as a function of $v_A$ in
Figure~\ref{fig:ene_scaling} (b). We find the acceleration rate normalization
$C \propto (v_A/v_{A0})^{1.36}$, where $v_{A0}$ is $300$ km/s in our normalization,
suggesting that the acceleration mechanism is a mixture of first-order Fermi mechanism
\red{
($\propto V/c$)~\citep[e.g.][]{Blandford1987Particle} and second-order Fermi mechanism
($\propto (V/c)^2$)~\citep{Fermi1949Origin}, where $V$ is the fluid speed ($\sim v_A$
in reconnection) and is typically much smaller than the light speed $c$.
This is because particles can gain energy in compression region and loss energy
in expansion region in the reconnection layer. If the compression region and
expansion region are uniformly distributed in the reconnection layer, particles
will experience a second-order Fermi acceleration similar as the original idea of
Fermi~\citep{Fermi1949Origin}. Instead, on average, particles
experience a net compression as they move into the reconnection layer, where plasma
is strongly compressed as shown in Figure~\ref{fig:mhd_fields}.
}
Since the reconnection layer is dynamically evolving,
the acceleration rate is time-dependent as well. Figure~\ref{fig:ene_scaling} (c)
and (d) show time evolution of the acceleration rate index $\alpha$ and the
acceleration rate normalization $C$. The $\alpha$ index fluctuates throughout the
simulation. For the three cases with stronger acceleration, the power-law index
fluctuates around 1.1. For the case with $v_A = 300$ km/s, the index is larger,
which is likely due to statistical errors as only a small number of particles
are accelerated to high energy. Figure~\ref{fig:ene_scaling} (d) shows the
acceleration rate generally decreases as the simulation evolves, which is likely
because reconnection becomes saturated in the late stage.

\subsection{Particle Acceleration due to Compression: Energy Dependent Diffusion Coefficients}
\begin{figure}[htbp]
  \centering
  \includegraphics[width=0.85\textwidth]{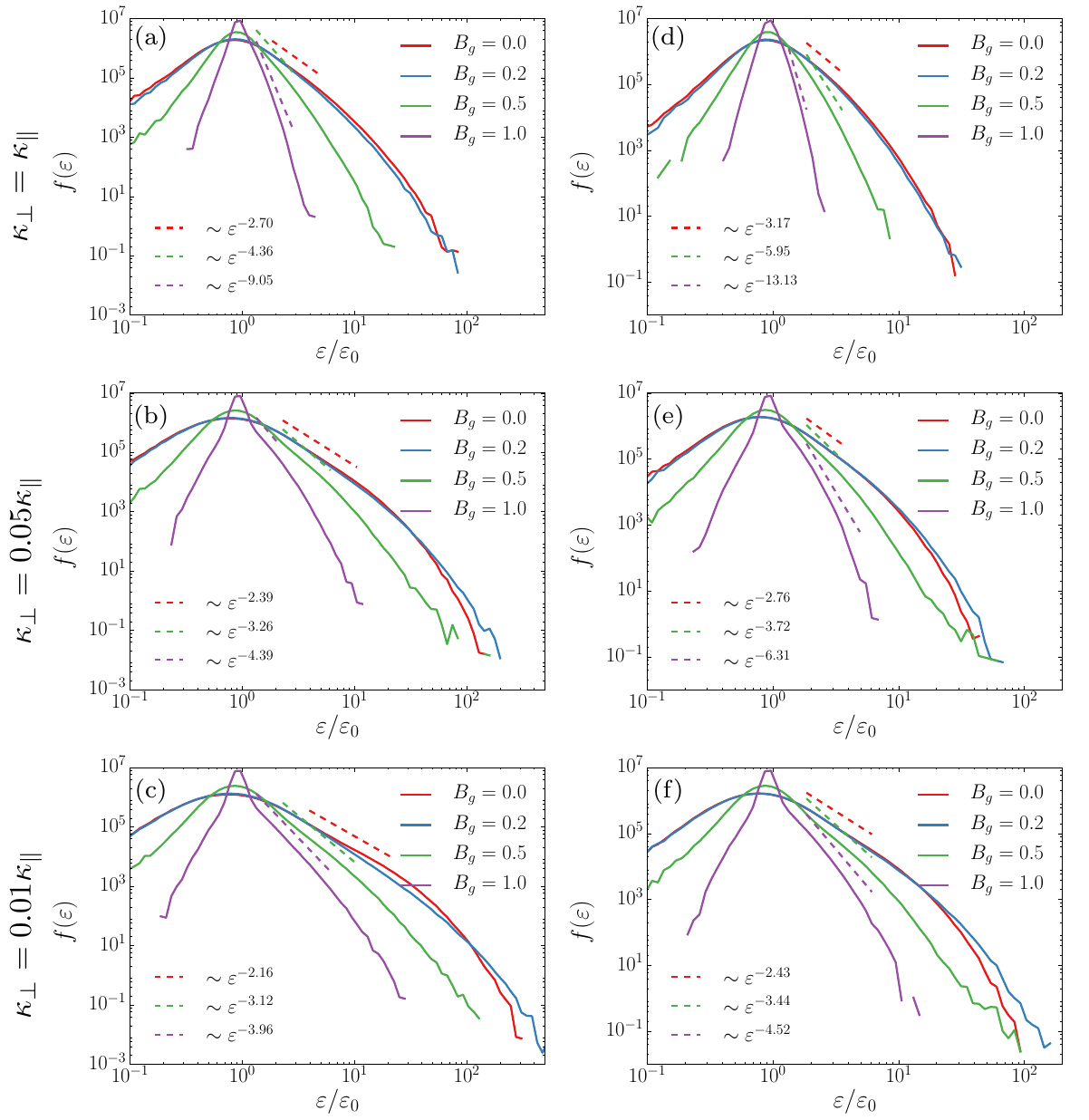}
  \caption{
    \label{fig:fe_kmodels}
    Particle energy distributions when $\kappa \propto p^{4/3}$ for cases with
    different $\kappa_\perp/\kappa_\parallel$.
    For (a)--(c), we assume that the particles are protons with an initial energy
    10 keV and an initial $\kappa_\parallel = 0.003\kappa_0$.
    For (d)--(f), we assume that the particles are electrons with an initial energy
    1 keV and an initial $\kappa_\parallel = 0.008\kappa_0$.
  }
\end{figure}

The constant and isotropic diffusion coefficient is a simplified assumption. In reality,
$\kappa$ usually depends on particle momentum. According to the quasi-linear theory
(Equation~\ref{equ:kpara_qlt}), $\kappa_\parallel\sim p^{4/3}$ for nonrelativistic
particles propagating in magnetic turbulence with a Kolmogorov power spectrum.
The diffusion coefficient in directions parallel and perpendicular to
the magnetic field can be quite different and previous test-particle calculations
give a perpendicular diffusion coefficient about a few percent of the parallel diffusion.
Figure~\ref{fig:fe_kmodels} shows the final energy spectra when we use
energy dependent $\kappa_\parallel = \kappa_1 (p/p_0)^{4/3}$
($\kappa_1 = 0.008\kappa_0$ for electrons and $0.003\kappa_0$ for protons) with  
3 different $\kappa_\perp / \kappa_\parallel$: $\kappa_\perp=\kappa_\parallel$,
$\kappa_\perp=0.05\kappa_\parallel$, and $\kappa_\perp=0.01\kappa_\parallel$.
The figure shows several trends. First, particles still develop power-law energy
spectra, but power-law energy range is shorter and the spectra roll over at certain
energies depending on the diffusion model.
The maximum particle energies are lower compared with the case with constant $\kappa$
because high-energy particles can escape from the acceleration regions much easier
due to their larger diffusion coefficients.
Second, as the ratio $\kappa_\perp/\kappa_\parallel$ decreases, the spectra become harder and
the maximum energy is higher. The spectra change dramatically for cases with
$B_g=1.0$. The power-law index $s$ changes from $s \sim 8.5$ to $s \sim 4$ for protons and
from $s \sim 12$ to $s \sim 4.5$. This is because when cross-field diffusion gets smaller,
particles could stay in the acceleration regions for a longer time.
Third, the maximum energies get close for cases with weak or moderate guide field
($B_g\leq 0.5$) even though the power-law part is steeper for cases with higher
guide field. Finally, in all the cases, protons can be accelerated to hundreds of
keV and electrons can be accelerated to tens of keV. For the case with
$\kappa_\perp=0.01\kappa_\parallel$, protons are accelerated to a few MeV and
electrons are accelerated to 100 keV, which are consistent with solar flare observations.

\begin{figure}[htbp]
  \centering
  \includegraphics[width=0.6\textwidth]{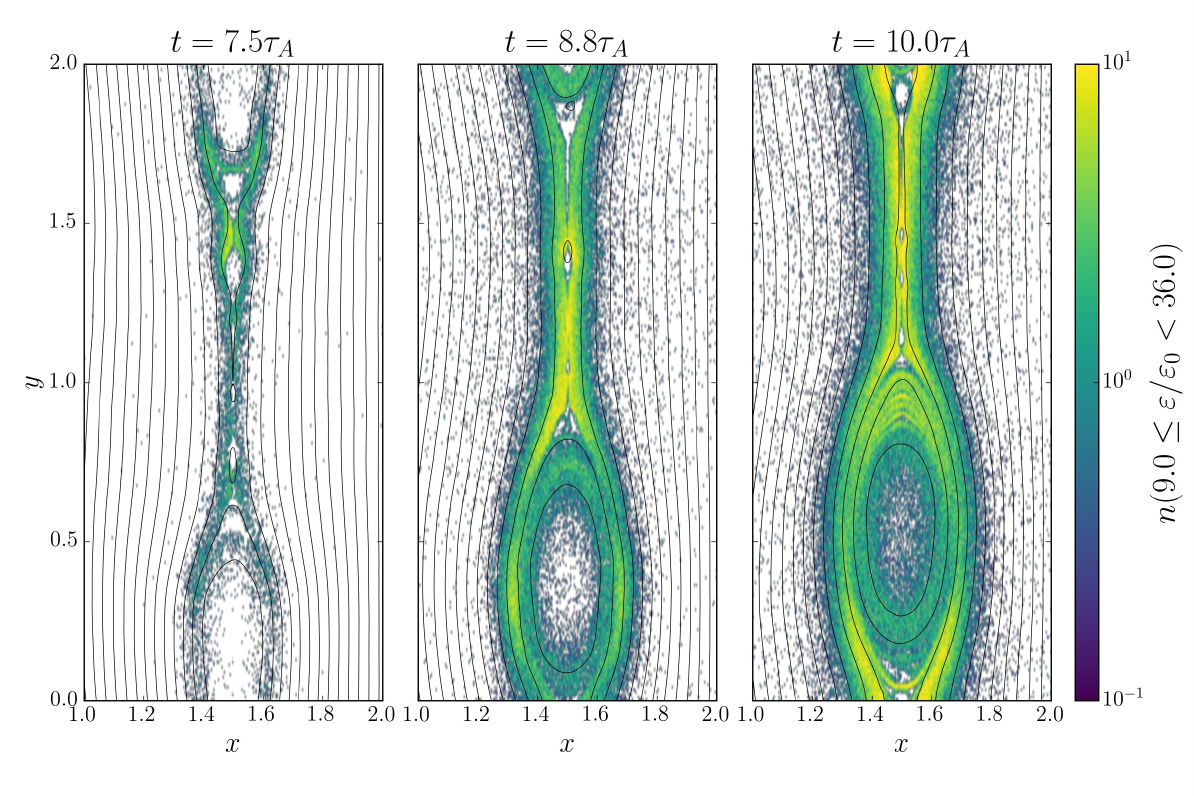}
  \caption{
    \label{fig:spatial_dist}
    Spatial distributions of high-energetic particles for the MHD run without a
    guide field at $t=7.5\tau_A$, $8.8\tau_A$ and $10.0\tau_A$.
    Here, we assume that the particles are electrons with an initial
    energy $\varepsilon_0=$1 keV, $\kappa_\parallel=0.008\kappa_0$, and
    $\kappa_\perp/\kappa_\parallel=0.01$. We choose particles with energy
    $9.0\leq\varepsilon/\varepsilon_0<36.0$.
  }
\end{figure}

The accelerated particles are not uniformly distributed in simulations.
Figure~\ref{fig:spatial_dist} shows the spatial distributions of high-energy
electrons ($9-36$ keV) for the simulation using the MHD run with $B_g=0$,
$\kappa_\parallel=0.008\kappa_0$, and $\kappa_\perp=0.01\kappa_\parallel$.
At an earlier time ($t=7.5\tau_A$), high-energy electrons are mostly in the island
at $y\sim 1.4$, the top side of the large island at $y\sim 0.5$, and the island merging
region at $y\sim 1.65$, suggesting that these regions are efficient at
accelerating particles. As the simulation evolves, high-energy particles are
advected with reconnection outflow and also diffuse to broader regions.
Close to the end of the simulation ($t=10\tau_A$), high-energy particles become
more uniform but their distribution still peaks at the two ends of the large
magnetic island and in the reconnection exhausts. This geometry is similar to the
above-the-loop-top hard X-ray sources observed in solar flares
\citep{Krucker2010Measure,Oka2015Electron}. The
confinement of high-energy electrons could potentially explain hard $X$-ray
emission  observed by \textit{RHESSI}.

\subsection{Trajectories of Pseudo-Particles}
\begin{figure}[htbp]
  \centering
  \includegraphics[width=0.95\textwidth]{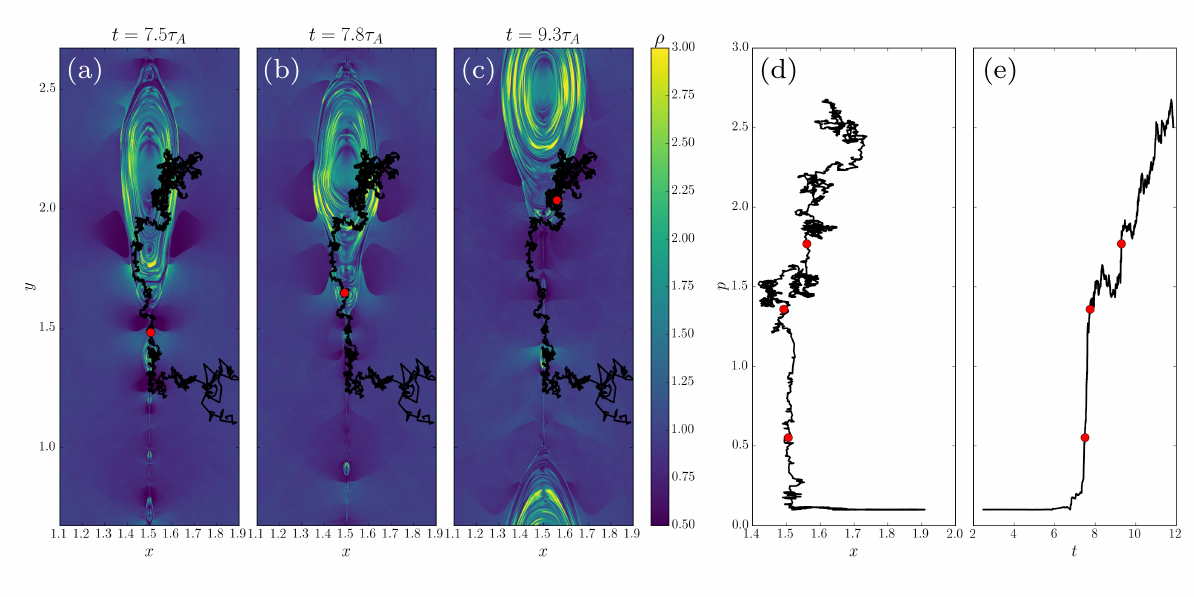}
  \caption{
    \label{fig:traj}
    Pseudo particle trajectory in the case with a constant $\kappa=0.003\kappa_0$
    and in the MHD simulation without a guide field. (a)--(c) show the trajectory
    with plasma density $\rho$ as background at three different time frames. The
    initial plasma density $\approx 1.0$.
    The red dots indicate the particle positions at each time frame. (d) particle
    momentum versus $x$ position. (e) particle momentum versus time. The three red
    dots indicate the three time frames shown in (a)--(c). The initial particle
    momentum is 0.1. Since we use periodic boundary conditions, We have
    shifted the background and particle trajectory when the particle crossed the
    boundary at $y=2.0$ for better visualization.
  }
\end{figure}

To further illustrate how particles are accelerated, Figure~\ref{fig:traj} shows a
representative pseudo-particle trajectory in the case with a constant
$\kappa=0.003\kappa_0$ and without a guide field.
We mark three red dots to indicate the
three major acceleration phases, including reconnection exhaust, contracting islands,
and island-merging regions. Initially, the particle slowly gets advected into the
reconnection layer. It gains energy in a short period of time ($7\tau_A<t<8\tau_A$)
when the particle diffuses across the reconnection current sheet, where the background
plasma is highly compressed.
\red{
  This indicates the particle acceleration in reconnection exhaust is dominated by
  a first-order Fermi process.
}
The particle is then trapped in a magnetic island and gains
more energy but the rate of energy increase becomes lower. This is because particles
can lose energy when they cross expanding regions of the magnetic island
(Figure~\ref{fig:mhd_fields}). In the late phase, the small island merges with
the large island and the particle gets accelerated and decelerated multiple times
but still gains more energy on average.
\red{
  These results indicate that particle acceleration in contracting islands and
  island-merging regions is a mixture of first-order and second-order Fermi processes
  but is dominated by the first-order process. This is due to the multiple
  compression and expansion layers in these regions and also the oscillations caused
  by merging magnetic islands. Note that the contracting island
  is a favorable region for the first-order Fermi acceleration but the pseudo
  particle trajectory shows that the contracting island (Figure~\ref{fig:traj} (b)
  and the middle red dot in (d) and (e)) is the not the dominant mechanism.
  Some other trajectories do show that the contracting island can be the
  dominant acceleration process (not shown here).
}

\section{Discussion and Conclusion}
\label{sec:con}
In this work, we have studied particle acceleration in a large-scale reconnection
site through solving the Parker energetic particle transport equation using velocity
and magnetic fields from high-Lundquist number MHD simulations of magnetic reconnection. We found
that compression in the reconnection layer leads to significant
particle acceleration and the formation of power-law energy distributions for both
electrons and ions. As the guide field becomes stronger, the power-law distribution
gets steeper, and the energy rollover of the power-law distribution and the maximum
particle energy decrease. The power-law index for electrons is about $2.4-13.1$,
depending on the guide field strength, which is close to the range found in solar
flare observations~\citep{Effenberger2017Hard, Oka2018Electron} and \red{the
observations of electron SEP events~\citep{Krucker2007Solar, Krucker2009Spectra}}.
The strong dependence
of particle acceleration on the guide field may be tested in observations (e.g.,
\citet{Qiu2010Reconnection}). When the perpendicular spatial diffusion is
much smaller than the parallel diffusion, we found the maximum electron energy
reaches $\sim 100$ keV and the maximum proton energy reaches a few MeV.
Detailed analysis shows that the acceleration rate $\propto v_A^{1.36}$, indicating
a mixture of first-order Fermi and second-order Fermi processes.
\red{
Pseudo particle trajectories show that the particle acceleration in reconnection exhaust
is dominated by first-order Fermi processes and that the acceleration in contracting
and merging magnetic islands is a mixture of first-order and second-order Fermi
processes but is still dominated by the first-order Fermi processes.
}

Our simulations also generate 2D spatial distributions of energetic particles.
We found the energetic particles are concentrated in reconnection exhausts and
magnetic islands. If combined with a radiation model, the 2D distributions could
be used to make predicted radiation map that is comparable with hard X-ray observation
by \textit{RHESSI} and \textit{FOXSI} and microwave imaging by radio observatories
such as Very Large Array (VLA) and Expanded Owens Valley Solar Array (EOVSA)
\citep{Gary2018Microwave}.

Our results are consistent with~\citet{Drury2012First}, which shows that the
spectral index depends on compressibility of the reconnection layer. But we found
that the spectral index is not just a simple expression of the compression ratio
between the outflow and inflow regions. This is likely due to the complex structures
(e.g. magnetic islands) and multiple compression and expansion regions
formed in the reconnection layer.
We found in our simulations that the particle energy spectra depend on the diffusion
model, especially the ratio of perpendicular diffusion coefficient and parallel
diffusion coefficient. Particle diffusion processes depend on the properties of
turbulence in the reconnection region such as turbulence
spectrum, the turbulence amplitude, the correlation length, and the turbulence
anisotropy, which are still under active research ~\citep{Huang2016Turbulent,
Beresnyak2017Three, Kowal2017Statistics, Loureiro2017Collisionless, Loureiro2017Role,
Mallet2017Disruption, Boldyrev2017MHD, Comisso2018MHD, Walker2018Influence,
Dong2018Role}. We expect a better understanding of these turbulence properties and
hence the particle diffusion processes in a reconnection layer in the near
future.

\red{
Our results are also consistent with in-situ observations in Earth's magnetotail.
Specifically, using the spacecraft measurements,~\citet{Fu2013Dipolar} found that
the reconnection layer is compressible and plasmoids are easily formed in this
compressible layer;~\citet{Fu2013Energetic} pointed out that the compressibility
of the reconnection layer can affect the contraction of magnetic islands and hence
the electron acceleration efficiency.
}

While fluid compression is the only acceleration mechanism considered in this study,
incompressible effects (e.g. fluid shear) could also accelerate
particles~\citep{Drake2006Electron, Zank2014Particle, LeRoux2015Kinetic, Li2018Roles},
potentially leading to stronger particle acceleration than that in observations.
Quantifying how other mechanisms change the particle spectral shape and maximum
energies may be important for future studies.

\red{
The developed numerical tools are not limited to study particle acceleration in
large solar flares. It can also be used to study particle acceleration at the
reconnection sites of nano flares, which have been proposed as a candidate
for explaining the power-law energy spectrum of superhalo electrons in solar
wind at quite times~\citep{Wang2012Quiet, Wang2015Solar}. We defer this to
a future work.
}

Our 2D simulations have a few limitations. First, the periodic boundary conditions
allows the large island to grows the system size, while in a solar flare, the largest
island is likely to be ejected out of the reconnection layer and cannot grow
to the system size, thus the current boundary conditions might lead to stronger particle
acceleration. Second, the 2D configuration prevents the field variation along the
out-of-plane direction, which might affect compression energization that depends
on the divergence of fluid velocity. Third, we use a plasma $\beta=0.1$ instead
of a lower plasma beta which may be present for solar flares, due to technical
difficulties when doing high-Lundquist-number simulations. Lower plasma $\beta$
might lead to stronger compression and hence stronger particle acceleration.

To conclude, we find that fluid compression in a reconnection layer leads to
significant particle acceleration and the formation of power-law energy
distributions for both electrons and ions. The compressibility of the reconnection
region, which depends on the guide field, determines the spectral index and cutoff
energy of the power-law distribution, and the maximum particle energy. The diffusion
coefficient and its anisotropy also influence the key features of the nonthermal
particle spectra. Our analysis shows that the acceleration in the reconnection layer
is a mixture of first-order Fermi and second-order Fermi processes. Our model
includes the acceleration mechanism derived from fully kinetic PIC simulations \citep{Li2018Roles},
and also applies to a macroscopic reconnection layer like in a solar flare. The resulting
time-dependent spatial and energy distributions of energetic particles can provide
explanations for observed energetic particle emissions in solar flares and other
astrophysical regimes.

\acknowledgments
This work was supported by NASA grant NNH16AC60I.
F.G. acknowledges the support in part from the National Science
Foundation under grant No. 1735414 and support from by the U.S.
Department of Energy, Office of Science, Office of Fusion
Energy Science, under Award Number DE-SC0018240. 
We also acknowledge the support by the DOE through the LDRD program at LANL.
Simulations were performed with LANL institutional computing.

\bibliography{references}{}
\bibliographystyle{aasjournal}
\end{document}